\documentclass[showpacs,amsmath,amssymb,aps,10pt,reprint,superscriptaddress,prl]{revtex4-1}
\usepackage{graphicx}
\usepackage{bm}
\usepackage[breaklinks=true,colorlinks=true,linkcolor=blue,urlcolor=blue,citecolor=blue]{hyperref}
\usepackage{graphics}
\usepackage{dcolumn}
\usepackage{amsmath,amssymb}
\usepackage{mathdots}
\usepackage{natbib}
\usepackage{soul,color}
\usepackage{epstopdf}
\usepackage{wasysym}

\usepackage{lipsum}

\begin{document}

\title{Reduction of microwave loss by mobile fluxons in grooved Nb films}

\author{O. V. Dobrovolskiy}
    \email{Dobrovolskiy@Physik.uni-frankfurt.de}
    \affiliation{Physikalisches Institut, Goethe University, 60438 Frankfurt am Main, Germany}
    \affiliation{Physics Department, V. Karazin National University, 61077 Kharkiv, Ukraine}
\author{R. Sachser}
    \affiliation{Physikalisches Institut, Goethe University, 60438 Frankfurt am Main, Germany}
\author{V. M. Bevz}
    \affiliation{Physics Department, V. Karazin National University, 61077 Kharkiv, Ukraine}
    \affiliation{ICST Faculty, Ukrainian State University of Railway Transport, Kharkiv, Ukraine}
\author{A.~Lara}
    \affiliation{Department of Condensed Matter Physics, Universidad Autonoma de Madrid, Spain}
\author{F.~G.~Aliev}
    \affiliation{Department of Condensed Matter Physics, Universidad Autonoma de Madrid, Spain}
\author{V. A. Shklovskij}
    \affiliation{Physics Department, V. Karazin National University, 61077 Kharkiv, Ukraine}
\author{A. I. Bezuglyj}
    \affiliation{Physics Department, V. Karazin National University, 61077 Kharkiv, Ukraine}
    \affiliation{Institute for Theoretical Physics, NSC-KIPT, 61108 Kharkiv, Ukraine}
\author{R. V. Vovk}
    \affiliation{Physics Department, V. Karazin National University, 61077 Kharkiv, Ukraine}
    \affiliation{ICST Faculty, Ukrainian State University of Railway Transport, Kharkiv, Ukraine}
\author{M. Huth}
    \affiliation{Physikalisches Institut, Goethe University, 60438 Frankfurt am Main, Germany}

\begin{abstract}
  In the mixed state of type II superconductors penetrated by an external magnetic field in the form of a lattice of Abrikosov vortices the dc resistance is known to increase with increasing velocity of the vortex lattice. Accordingly, vortex pinning sites impeding the vortex motion are widely used to preserve the low-dissipative response of the system. Here, while subjecting superconducting Nb films with nanogrooves to a combination of dc and ac current stimuli and tuning the number of mobile and pinned vortices by varying the magnetic field around the so-called matching values, we observe a completely opposite effect. Namely, the vortex-related microwave excess loss for mobile vortices becomes smaller than for pinned vortices in a certain range of power levels at ac current frequencies above 100\,MHz. While a theoretical description of the observed effect is yet to be elaborated we interpret our findings in terms of a competition of the effective cooling of the system by the quasiparticles leaving the vortex cores with the conventional Joule heating caused by the current flow.
\end{abstract}

\maketitle   

\section{Introduction}

Superconductivity is a macroscopic quantum coherent state observed in most materials when they are cooled below their critical temperature $T_c$. The superconducting state is destroyed by large magnetic fields $H$ or transport currents $I$ when their values exceed the critical field $H_{c2}$ and the depairing current $I_c$, respectively. Interestingly, at sufficiently high-power levels an electromagnetic field of subgap frequency, which is of the order of a few hundred GHz for a superconductor with a $T_c$ of $5$\,K, may stimulate superconductivity itself. Discovered in 1966 by Wyatt \emph{et al} \cite{Wya66prl} via an enhancement of the supercurrent in tin films in the Meissner state in the presence of a microwave (mw) electromagnetic field, this counterintuitive non-equilibrium effect has inspired extensive research both experimentally \cite{Pal79prb,Tol83etp,Hes93prb,Pal80prl,Ent81prb,Bec13prl,Vis14prl} and theoretically \cite{Eli70etp,Asl78etp,Asl82etp,Lem83etp}. A comprehensive review can be found in Ref. \cite{Zol13ltp}.

Stimulation of superconductivity at mw frequencies was observed in various type-I superconducting systems such as thin films \cite{Pal79prb,Tol83etp}, cylinders \cite{Pal80prl} and almost all types of weak links \cite{Hes93prb}. In spatially homogeneous superconductors, the effect was explained by Eliashberg in 1970 to be a consequence of an irradiation-induced redistribution of quasiparticles away from the gap edge \cite{Eli70etp}. Later on, Aslamazov and Larkin studied mw-stimulated superconductivity in a contact between superconductors and showed that an increase of the critical current may result as a consequence of the effective cooling of electrons trapped in the region of the contact which has a lower value of the superconducting gap $\Delta$ \cite{Asl78etp}. Recently, mw-stimulated superconductivity has been observed in transient regimes in NbN films \cite{Bec13prl} and was demonstrated to improve the quality factor of superconducting microwave resonators \cite{Vis14prl}.

When shaped in technology-relevant geometries, such as thin films \cite{Dob15met}, nanowires \cite{Kom14apl,Win14apl,Sen15apl}, thin-wall cylinders \cite{Tsi16pcm} and nanotubes \cite{Fom12nlt,Rez14pcs,Cor18nal}, almost all superconductors fall into the class of type-II superconductors. In external magnetic fields, whose magnitude is between the lower and the upper critical field, these are penetrated by a lattice of Abrikosov vortices and the vortex dynamics determines their resistive response. In addition to two-dimensional structures fabricated by conventional thin films techniques, superconducting nanostructures can be fabricated by strain engineering \cite{Thu10nlt} or direct-write techniques such as focused ion \cite{Kom14apl,Cor18nal,Ser16bjn} and electron \cite{Win14apl,Sen15apl,Hut18mee,Mak14nsr} beam-induced deposition. The latter technique is particularly suitable for the fabrication of three-dimensional structures and nano-architectures with non-trivial topology, such as a torus, a helix, a M\"obius band and so on. We are aware of the availability of such superconducting architectures in the nearest future and anticipate their exploitation in miniature electronic devices operating at high and ultra high ac current frequencies.

Surprisingly, while superconducting circuit elements are used in various applications operated at frequencies in the radiofrequency and mw range, such as single photon detectors \cite{Mar13nph}, resonators \cite{Vis14prl}, filters \cite{Leh17sst,Dob15apl}, transistors \cite{Vla16nsr}, generators \cite{Dob18apl}, transmission lines for circuit quantum electrodynamics \cite{Wal04nat,Hof09nat} and quantum information processing \cite{Dic09nat}, the mw stimulation effect in the vortex state has largely been remained unexplored so far. In the only available work \cite{Lar15nsr}, superconductivity was stimulated by an in-plane field component generated by an mw current flowing through a coplanar waveguide placed near the surface of Pb films. The stimulation effect was more clearly seen in the enhancement of $H_{c2}$ and less pronounced in the increase of $T_c$ \cite{Lar15nsr}. The question remained whether there is a stimulation effect when a high-frequency current flows through the superconductor itself. This question is addressed in this Letter. Specifically, we study mw-stimulated superconductivity in Nb films with nanogrooves inducing a periodic pinning potential of the washboard type \cite{Dob17pcs} for the vortex motion across the grooves. As the ac power increases, in the vortex-related mw loss we observe a crossover from minima at low frequencies to maxima at high frequencies. This means that superconductivity is stimulated when the fluxons are \emph{not pinned but are mobile}, that is the opposite effect to the one known for the vortex-flow dc dissipation.

\section{Experimental}

The sample is a 150\,$\mu$m-wide microstrip fabricated by lithography from a 40\,nm-thick epitaxial Nb film. The Nb films was sputtered by dc magnetron sputtering on an a-cut sapphire substrate in a setup with a base pressure in the 10$^{-8}$\,mbar range. In the sputtering process the substrate temperature was 850$^\circ$C, the Ar pressure 4$\times$10$^{-3}$\,mbar, and the growth rate was about 1\,nm/s. X-ray diffraction measurements revealed the (110) orientation of the film \cite{Dob12tsf}. The epitaxy of the films has been confirmed by reflection high-energy electron diffraction. The as-grown films have a smooth surface with an rms surface roughness of less than 0.5\,nm, as deduced from atomic force microscopy scans in the range $1\times1\,\mu$m$^2$. The superconducting transition temperature $T_c$ of the sample is $8.66$\,K. The upper critical field of the Nb film at zero temperature $H_{c2}(0)$ is about $1$\,T as deduced from fitting the dependence $H_{c2}(T)$ to the phenomenological law $H_{c2}(T) = H_{c2}(0)[1 -(T/T_c)^{1/2}]$.
\begin{figure}[t]
    \centering
    \includegraphics*[width=0.8\linewidth]{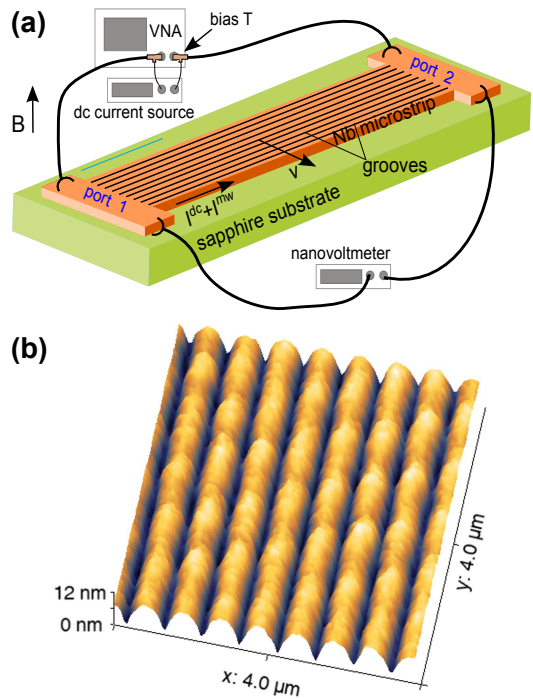}
    \caption{(a) Experimental geometry. A combination of dc and mw currents is applied to the superconducting Nb microstrip. The microstrip contains an array of 500\,nm-spaced nanogrooves milled by Ga-FIB. The Lorentz force causes the vortices to move against the barriers of the periodic pinning potential induced by the nanogroove array. (b) Atomic force microscopy image of the surface of the Nb film with an array of nanogrooves.}
    \label{f1}
\end{figure}

An array of 10\,nm-deep, 50\,nm-wide and 500\,nm-spaced nanogrooves was milled by focused ion beam (FIB) in a high-resolution dual-beam microscope (FEI, Nova NanoLab 600). The grooves are parallel to the long side of the microstrip with a misalignment of less than 0.2$^\circ$, such that the vortices move across the grooves under the action of the transport current, Fig. \ref{f1}. In the FIB milling process the beam parameters were 30\,kV/50\,pA, 1\,$\mu$s dwell time and 50\,nm pitch. The microstrip width is an integer multiple number ($N=$300) of the nanopattern period to prevent possible ratchet effects due to the edge barrier asymmetry \cite{Pry06apl}. The background pinning in the epitaxial (110) Nb films is very weak as compared to the anisotropic pinning induced by the nanogrooves \cite{Dob12njp}.
\begin{figure*}[htb]%
    \includegraphics[width=0.85\linewidth]{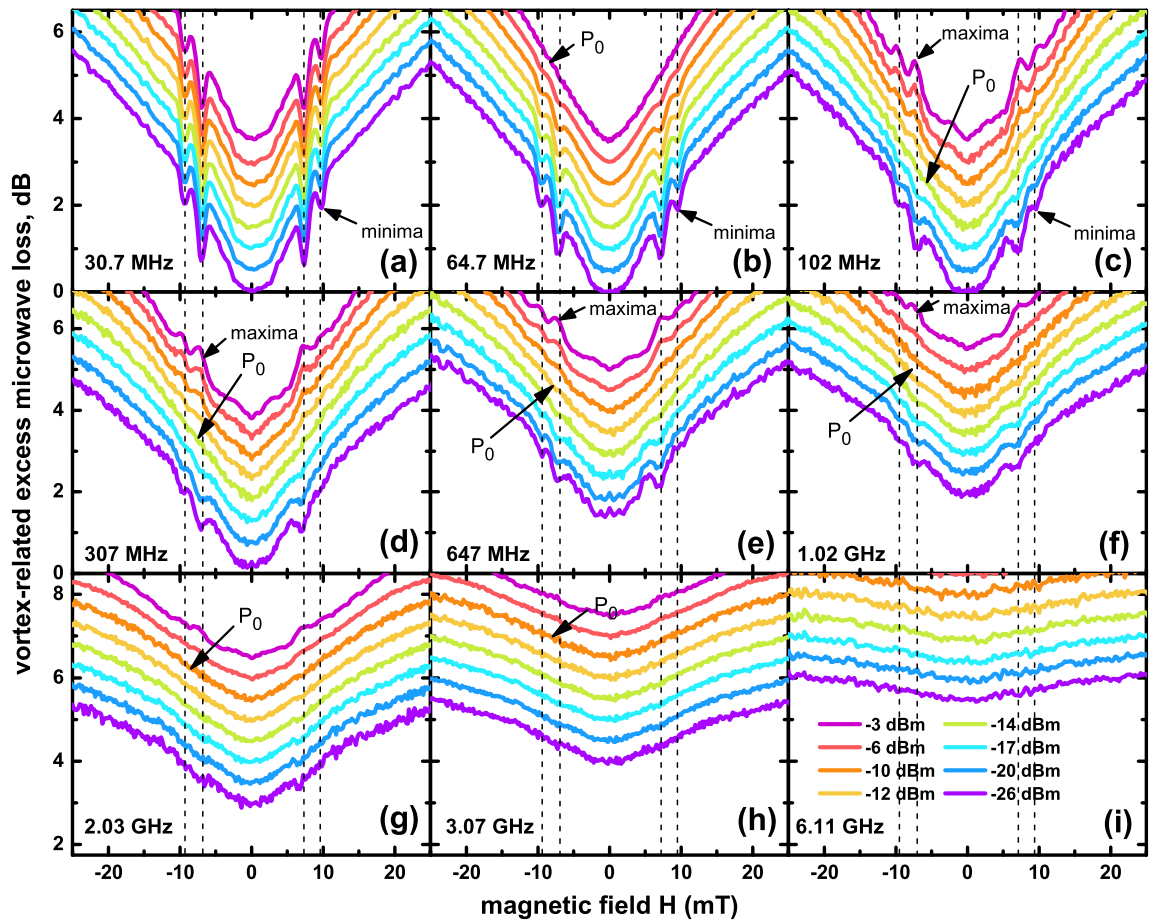}
  \caption[]{Vortex-related mw excess loss as a function of the magnetic field at $T = 0.9T_c$ for a series of ac frequencies and mw power levels, as indicated. An offset of $+0.5$\,dB for each subsequent curve is used to facilitate reading the data. The vertical dashed lines indicate the matching fields $H_{m1}$ = $7.2\,$mT and $H_{m2}$ = $9.6$\,mT at which the vortex lattice is commensurate with the pinning nanostructure. The optimal microwave power $P_0$, at which excess loss minima turn into maxima, is indicated. The stimulation effect is seen in panels (b) to (g). The minimum at $H=0$ is shifted to larger loss values from (a) to (i) due to the delocalization transition in the vortex dynamics at the depinning frequency.}
    \label{f2}
\end{figure*}

The microwave vortex-state response was studied by cryogenic broadband spectroscopy \cite{Dob15apl} with magnetic field $H$ directed perpendicular to the film surface. The mw signal was generated and detected by an Agilent E5071C vector network analyzer (VNA). The mw and dc signals were superimposed and uncoupled by using two bias-tees mounted at the VNA ports. In what follows we only present the vortex-related contribution in the mw loss, as deduced from the relative change of the absolute value of the forward transmission coefficient $\Delta S_{21} \equiv S_{21}(f,P,H)/S_{21}(f,P,H_{\mathrm{ref}})$ where $S_{21}(f,P,H_\mathrm{{ref}})$ is the frequency- and mw-power-dependent reference loss at $H_\mathrm{ref} >H_{c2}$ \cite{Lar15nsr,Sol14prb}.

\section{Results and discussion}

Figure \ref{f2} shows the vortex-related excess mw loss as a function of the magnetic field at $T = 0.9T_c$ for a series of ac frequencies. The mw loss is maximal at higher $H$ values as the number of vortices in the sample increases and it is minimal at $H = 0$, as expected. On the background of the smooth crossover from the minimal to the maximal mw loss value, one clearly recognizes spikes at the magnetic fields $H_{m1} = 7.2$\,mT and $H_{m2}=9.6$\,mT in panels (a) to (g) of Fig. \ref{f2}. The spike character evolves with increasing frequency and the spikes eventually vanish above $2$\,GHz. At $H_{m1}$ and $H_{m2}$ the vortex lattice is commensurate with the 500\,nm-periodic array of nanogrooves such that all vortices are pinned at the groove bottoms at 7.2\,mT, which is referred to as a fundamental matching field. At $H_{m2}=9.6$\,mT one half of vortices is pinned at the groove bottoms while the rest are pinned between the grooves, so that this field value is referred to as a secondary matching field \cite{Dob15apl}. The grooves are known to act as strong pinning sites for the vortex motion across them \cite{Dob12njp,Ser16bjn}. Accordingly, the pinning of vortices is most efficient at $H_{m1}$, less so at $H_{m2}$ and even less efficient at an arbitrary value of $H$ between $H_{m1}$ and $H_{m2}$, say $H = 8$\,mT for definiteness. In the limit of low frequencies in Fig. \ref{f2}(a), which corresponds to the quasistatic regime, one observes minima in the vortex-related mw excess loss at both matching fields for all mw power levels in the accessible range from -26\,dBm to -3\,dBm (2.5\,$\mu$W to 0.5\,mW). Such matching dips are typically seen in the field dependence of the electrical dc resistance \cite{Ser16bjn,Dob12njp} or in the ac susceptibility in the linear regime at low mw power levels \cite{Sol14prb,Dob15apl,Lar15nsr}.

With increase of the frequency to 64.7\,MHz in Fig. \ref{f2}(b), at higher mw power levels, the minima become shallower and turn into maxima at $102$\,MHz in Fig. \ref{f2}(c). This crossover from minima to maxima at $H_{m1}$ and $H_{m2}$ is also seen in Fig. \ref{f2}(d) to (g) with further increase of the frequency. It continues as long as the field dependence of the vortex-related mw excess loss does not become flattened at frequencies above the depinning frequency $f_d$ \cite{Git66prl,Cof91prl,Shk11prb} and the data become more noisy at yet higher frequencies. The depinning frequency has the physical meaning of the crossover frequency from the low-frequency regime, where the pinning forces dominate and the vortex response is weakly dissipative, to the high-frequency regime, where the frictional forces prevail and the response is strongly dissipative \cite{Pom08prb,Sil17psr}. The depinning frequency depends on the temperature $f_d\equiv(T,H,I) = f_d(0,H,I)[1 - (T/T_{c})^4]$ \cite{Zai03prb}, magnetic field $f_d = f_d(T,0,I)[1 - (H/H_{c2})^2]$ \cite{Jan06prb}, and dc bias current $f_d = f_d(T,H,0)[1- (I/I_d)^2]^{1/2}$ \cite{Dob17nsr}. The zero-temperature value for our sample is $f_d \approx 5.72$\,GHz, as deduced from fitting the data acquired at $T = 0.3T_c$ \cite{Dob17nsr} to the expression $f_d\equiv(T,H,I) = f_d(0,H,I)[1 - (T/T_{c})^4]$, yielding $f_d (0.9T_c) \approx 1.96$\,GHz at small fields and currents. This correlates reasonably well with the disappearance of the matching peculiarities in Fig. \ref{f2}(g) as the pinning potential becomes less efficient for the vortex dynamics at higher frequencies. In all, the presence of the nanogroove array characterized by a strong washboard pinning potential allows for a direct comparison of the vortex-related mw loss for mobile and pinned vortices. However, this approach is limited to the frequency range below the depinning frequency, while the mw stimulation effect has been checked to continue up to the highest frequency accessible with our VNA (14\,GHz), as inferred from the enhancement of $H_{c2}$ and $T_c$ following the analysis of Ref. \cite{Lar15nsr}. Importantly, no saturation of the stimulation effect at 14\,GHz has yet been observed.

\begin{figure}[t]
    \centering
    \includegraphics[width=1\linewidth]{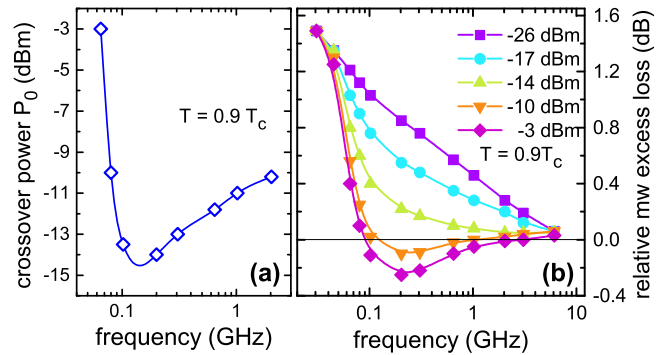}
    \caption{(a) Dependence of the optimal power $P_0$ on the mw power at $H_{m1} = 7.2\,$mT and $T = 0.9T_c$. (b) Frequency dependence of the relative insertion loss illustrating the frequency and mw power range of mw-stimulated superconductivity.}
    \label{f3}
\end{figure}

For a quantitative analysis of the excess mw loss due to mobile and pinned fluxons we introduce the crossover power $P_0$ corresponding to the crossover from the matching dips to peaks in Fig. \ref{f2}. The evolution of $P_0$ as a function of the ac current frequency is shown in Fig. \ref{f3}(a). At lower frequencies in the quasistatic regime there is no crossover from dips to peaks so that $P_0$ is undefined. The matching minima first vanish at $64.7$\,MHz at a relatively high mw power of $-3$\,dBm. The stimulation effect becomes more efficient at higher frequencies such that $P_0$ decreases to about $-14$\,dBm and attains a minimum at about $150$\,MHz. The further gradual increase of $P_0$ is associated with the approach of the depinning frequency $f_d$, above which $P_0$ becomes undefined because of the vanishing difference in the contributions of pinned and mobile vortices.

The relative mw excess loss between the mobile and pinned fluxons is shown in Fig. \ref{f3}(b). Positive loss values mean that pinned vortices dissipate less than the mobile ones while negative values result when the mobile vortices dissipate less than the pinned ones. The gradually decreasing loss difference at low mw power levels off with an increase of the frequency and reflects the delocalization transition in the vortex dynamics associated with the depinning frequency. Here, the relative loss is zero not exactly at $f_d$ but at about 8\,GHz due to the smoothed functional shape of the crossover from the low-dissipative response at low frequencies to the strongly dissipative response at high frequencies \cite{Dob15apl}. As is apparent from Fig. \ref{f3}(b), there is no stimulation effect in the whole frequency range unless the mw power is larger than about $-12$\,dBm. Namely, as soon as the mw power exceeds the $-10$\,dBm level the stimulation effect occurs between $0.1$ and $1$\,GHz and its magnitude becomes larger at $-3$\,dBm. This is accompanied by a broadening of the frequency range up to $2$\,GHz where the mw loss due to mobile vortices is smaller than that due to the pinned ones.

While no upper frequency limit for the effect has been observed below 14\,GHz, we have checked that the stimulation effect is suppressed as the mw power is further increasing to +6\,dBm. This allows us to introduce an additional power parameter $P_{\mathrm{sup}}$, at which a crossover from the stimulation to the suppression of superconductivity takes place. An extensive analysis of the stimulation effect on the thermodynamic ($T_c$, $H_{c2}$) and transport properties ($I_c$) will be published elsewhere. Here we just note that $P_{\mathrm{sup}}$ is related to the Joule heating and the stimulation of superconductivity between $P_0$ and $P_{\mathrm{sup}}$ is likely due to the effective cooling of the superconductor, which prevails over Joule heating caused by the current flow. Namely, under the action of the ac current of sufficiently high frequencies and power levels between $P_0$ and $P_{\mathrm{sup}}$ the quasiparticles leaving the vortex cores relax in an effectively larger superconducting volume due to the vortices shaken by the mw stimulus.

\section{Conclusion} By comparing the vortex-related excess mw loss for mobile and pinned fluxons we have observed a crossover from mw loss minima at low frequencies to the regime of mw-stimulated vortex-state superconductivity above $100$\,MHz in a certain range of mw power levels. The spatially periodic suppression of the superconducting gap in the vortex cores and the vortex dynamics under the action of the combination of dc and mw transport currents requires to simultaneously account for the kinetics of quasiparticles in conjunction with the spatial and temporal changes of the order parameter. While a theoretical description of the observed effect is yet to be elaborated we attribute our findings to a competition of the effective cooling of the system by quasiparticles leaving the vortex cores with the conventional Joule heating associated with the current flow.

\section*{Acknowledgements}
OD thanks the German Research Foundation (DFG) for support through Grant No 374052683 (DO1511/3-1). AL and FA acknowledge support by Spanish MINECO (MAT2015-66000-P) and Comunidad de Madrid (NANOFRONTMAG-CM S2013/MIT-2850) Grants. This work was supported by the European Cooperation in Science and Technology by via COST Action CA16218 (NANOCOHYBRI). Further, funding from the European Commission in the framework of the program Marie Sklodowska-Curie Actions --- Research and Innovation Staff Exchange (MSCA-RISE) under Grant Agreement No. 644348 (MagIC) is acknowledged.
\vspace*{-3mm}


%

\end{document}